\documentclass[showpacs, twocolumn]{revtex4-1}
\usepackage{graphicx}
\usepackage{amsmath}

\begin{document}

\title{Quantum gas of polar molecules ensembles at ultralow temperatures: $f$-wave superfluids} 

\author{Abdel\^{a}ali Boudjem\^{a}a}

\affiliation{Laboratory of mechanics and energy, Hassiba Benbouali University of Chlef P.O. Box 151, 02000, Chlef, Algeria.}

\begin{abstract}
We investigate novel $f$-wave superfluids of fermionic polar molecules in a two-dimensional bilayer system with dipole moments polarized perpendicular to the layers
and in opposite directions in different layers.
The solution of the BCS gap equation reveals that these unconventional superfluids emerge at temperatures on the level of femtokelvin
which opens up new possibilities to explore the topological $f+i f$ phase, quantum interferometry and Majorana fermions in experiments with ultracold polar molecules.
The experimental realization of such interesting novel $f$-wave pairings is discussed.

\end{abstract}


\maketitle

\section{Introduction}

The quest for ultralow temperatures has been a major subject of physics for over a century 
offering intriguing perspectives to observe new quantum phenomena that occur on very low energy scales.  
Ultracold temperature systems lead to vast improvements in precision measurements by allowing better atomic clocks and sensors. 
They conduct us also to test the fundamental laws of physics.
The liquefaction of helium at temperatures below 1K led to the discovery of superconductivity \cite {Onnes} and superfluidity \cite{Kapz}.
Recently, the development of laser and evaporative cooling techniques \cite{Chu, Tann, Bill} has paved the way to achieve Bose-Einstein condensates (BEC) and degenerate Fermi gases
at temperatures of the order of nanokelvin \cite{Cornel, Kitt2, Jin1}.
Nowadays gravito-magnetic traps are used to cool atoms down to even lower temperatures and reaching the picokelvin range \cite{Kitt, Lu, Tian}.

In this paper, we propose a novel scheme to achieve temperature well below any temperature that has been reached so far. 
We study, for the first time to our knowledge,  the formation of an unconventional $f$-wave superfluid of fermionic polar molecules 
in a two-dimensional (2D) bilayer geometry those dipole moments are oriented perpendicularly to the layers and in opposite directions in different layers.
In such a bilayer system only higher partial wave ($l \geq 1$) superfluids are allowed \cite{Fedo, Boudj} in contrast to what has been obtained in the setup
of Refs \cite{Demler, Pikov,Baranov1, Zin} where  an interlayer $s$-wave superfluid is formed due to the $s$-wave dipolar interaction between different layers.


Evidence of $f$-wave pairing was found first in ${}^3$He superluids nearly three decades ago \cite{Sauls}.
Mao et \textit{al}.\cite {Mao} have demonstrated that a chiral ($f+if$)-wave superconducting pairing may be induced in the lowest heavy hole
band of a hole-doped semiconductor thin film through proximity contact with an $s$-wave superconductor.
Cold atom physics provides also an ideal platform to realize and investigate such unconventional $f$-wave pairing.
For example, Lee et \textit{al}. \cite {Sarma} have pointed out that $f$-wave pairing superfluidity of spinless fermions atoms can emerge in $p_{x,y}$-orbital bands 
of 2D honeycomb optical lattices.
In the same context, Hung et {\textit al}. \cite{Hsian} found that frustrated Cooper pairing in the $p$-orbital bands in triangular lattice  with ultracold spinless fermions exhibits an
unconventional supersolid state with the $f$-wave symmetry.
Likewise, a chiral $f$-wave topological superfluid can be induced in cold fermionic-atom optical lattices through the laser-field-generated effective non-Abelian
gauge field, controllable Zeeman fields and s-wave Feshbach resonance \cite {Ning}. 

The realization  of  quantum degenerate gas of polar molecule which establish a long-range and anisotropic interaction among molecules,
opens wide avenues for modeling of condensed matter systems \cite{Baranov, Pfau,Carr, Pupillo2012}. 
Molecules are promising candidates for studying unconventional superfluids, as they possess tunable electric dipole moment 
which can be induced by a static dc electric field, in addition to their own intrinsic dipole moment  \cite{KK, Aik, Deig, Kev}. 

We consider two clouds of 2D fermionic polar molecules separated by a distance $\lambda$ much larger than the conffinement length
of the molecules within each layer. The dipole moments of polar molecules are aligned perpendicularly to the layers 
and in opposite directions in different layers. The orientation of the dipole moments can be controlled with dc and ac electric fields \cite{GCZ}.
Our analysis is based on many-body perturbative theory able to describe the renormalized  gap equation in the weak coupling regime by accounting 
corrections related to the effective mass of the quasiparticles. 
This enables us to derive a useful relation for the transition temperature $T_c$.
It is then found that $T_c$ of the interlayer $f$-wave superfluid is on the level of femtokelvins which constitutes a new way to create 
ultracold polar molecular gases in the limit of femtokelvin.
Most recent theoretical studies have revealed that $p$-and $d$-wave superfluids 
of fermionic polar molecules occur respectively, at temperatures on the order of nanokelvin and picokelvin \cite{Fedo, Boudj}.

In the weak coupling regime, the transition into the superfluid phase is of the Kosterlitz-Thouless scenario.
However, as was shown by Miyake \cite {Miy}, the Kosterlitz-Thouless transition temperature is very close to the critical temperature $T_c$ calculated within the BCS approach.
Therefore, our many-body perturbative theory offers reliable predictions for the critical temperature in the regime of interest.
We determine in addition, the value of the interlayer spacing, which would enable observing these novel superfluids.

The remainder of the paper is structured as follows. In Sec.\ref{Mod}, we present the model that describes the bilayer pairings
and investigate the $f$-wave scattering properties for the interlayer problem.
In Sec.\ref {TT}, we provide analytical expressions for the order parameter and the transition temperature.
We discuss, on the other hand, the possibilities to realize and detect the predicted polar molecules ensembles.
Conclusions are given in Sec.\ref{Conc}.

\section{The model } \label{Mod}

Assuming vanishing hopping between layers, the interaction potential between two molecules belonging to different layers is given by \cite{Fedo, Boudj}
\begin{equation} \label{pot} 
V(r) = - d^2\frac{r^2-2\lambda^2}{(r^2+\lambda^2)^{5/2}}.
\end{equation}
Here $- d^2$ is the scalar product of the average dipole moments of  the molecules.
The potential $V(r)$ is repulsive for $r < \sqrt{2} \lambda$, while it is attractive at large distance $r$ leading to the formation of an interlayer bound state (see Fig. \ref{FT}.a).
The existence of such a bound state signals  the possibility of the interlayer BCS pairing when the size of the interlayer bound state 
is much larger than the interparticle separation.
The Fourier transform of the potential (\ref{pot}) reads
\begin{align} \label{FT}
& V({\bf q}) = \int d{\bf r} V ({\bf r}) e^{- i {\bf q r}} = \frac{2\pi\hbar^2} {m} r_*|{\bf q}| e^{-|{\bf q}|\lambda}, 
\end{align}
where $r_*=md^2/\hbar^2$ is the characteristic dipole-dipole distance and $m$ is the particle mass. 
For  $q\lambda \ll 1$,  $V({\bf q}) = (2\pi\hbar^2/m) r_* q$.  This linear dependence on $q$  originates from the so-called anomalous contribution to scattering \cite{Zhen}. 
In the limit $q \rightarrow 0$, $V({\bf q}) \rightarrow 0$ (see Fig. \ref{FT}.b). Thus, the interparticle scattering will be dominated by higher-order corrections \cite{Baranov1}.



It is instructive then to solve the gap equation by calculating the $f$-wave scattering amplitude up to the second order following the exact method described in \cite{Fedo,Boudj}.\\
The off-shell $f$-wave scattering amplitude is given by 
\begin{equation}\label{scat}
f_3 (k',k)=\int_0^{\infty}  J_3 (k'r) V(r) \psi _3(k,r)  2\pi r dr,  
\end{equation}
where $J_3$ is the Bessel function and $\psi_3 (k,r)$ is the true wavefunction of the $f$-wave relative motion with momentum $k$. 
It is normalized in such a way that for $r \rightarrow \infty $, one has $\psi_3 (k,r)=J_3 (kr)- i (m/4\hbar^2) f_3(k)  H_3 (kr)$, with $H_3$ being the Hankel function.
Then, the  off-shell scattering amplitude (\ref{scat})  reads \cite{Boudj}
\begin{equation}\label{scat1}
f_3 (k',k)=\frac{\bar f_3(k',k)}{1+i  (m/4\hbar^2) \bar f_3(k)}.
\end{equation}
Here $\bar f_3$ is real and represented as a sum of two terms: $\bar f_3=\bar f_3^{(1)}+\bar f_3^{(2)}$, 
where $\bar f_3^{(1)}$ and $\bar f_3^{(2)}$ stand for the first and second order corrections to the off-shell scattering amplitude, respectively. 

For $k_F r_* \ll 1$, the $f$-wave channel of the first order contribution to the off-shell scattering amplitude can be evaluated in the Born approximation as \cite{Boudj}
$\bar f_3^{(1)} (k',k)= \int \frac{ d\phi_{k'} d\phi_k}{(2\pi)^2} e^{3i(\phi_{k'}-\phi_k)} V ({\bf r}) e^{i {\bf (k-k') r}} =(2\pi\hbar^2 k r_*/m) {\cal F}_3^{(1)} (k',k)$, where

\begin{figure}
\includegraphics[width=8.5 cm,height=5.5 cm,clip]{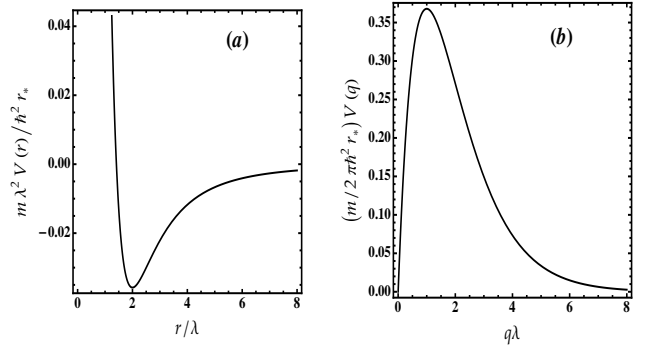}
\caption {The interlayer potential $V(r)$ (panel a) and its Fourier transform $V(q)$ (panel b).}
  \label{FT}
\end{figure}

\begin{align}\label{scat2}
{\cal F}_3^{(1)} (k'\lambda,k\lambda)& = \frac{1}{k \lambda} \int_0^{\infty} x d x J_3 (k' \lambda x) J_3 (k \lambda x) \nonumber \\
& \times \frac{x^2-2}{(x^2+1)^{5/2}}.
\end{align} 
At $k=k_F$, ${\cal F}_3^{(1)} (k_F\lambda)$ takes the form
\begin{align}\label{scat2}
{\cal F}_3^{(1)} (k_F\lambda)&=  \left [\frac{30}{ (k_F\lambda)^2}+2\right] L_3(2 k_F\lambda) - \frac{112 (k_F\lambda)^2-12} {105 \pi }\nonumber \\
&-\left [6 (k_F\lambda)^5 \, _0\tilde{F}_1(;7;(k_F\lambda)^2)+2I_7 (2k_F\lambda)\right] \nonumber \\
  & -\frac{2 \left[7 (k_F\lambda)^2+60\right] L_4(2 k_F\lambda)}{(k_F\lambda)^3} , 
\end{align} 
where $_0\tilde{F}_1$ is the regularized confluent hypergeometric function, $L$ is the Struve function 
and $I$ is the modified Bessel function of the first kind. For $k_F\lambda \ll 1$, ${\cal F}_3^{(1)} (k_F\lambda) \approx 4/(35 \pi) $. 
Equation (\ref{scat2}) shows that $\bar f_3^{(1)}$  is negative for $k_F \lambda \leq 3$ signaling that the system undergoes a transition
into an interlayer $f$-wave pairing at extremely low temperatures.

For the second order correction, only the on-shell $f$-wave form is important. It is given by 
$\bar f_3^{(2)} (k)=(2\pi^2\hbar^2 k^2 r_*^2/m)  {\cal F}_3^{(2)}  (k)$, where
\begin{align}\label{scat3}
{\cal F}_3^{(2)} (k\lambda)& =\frac{1}{(k \lambda)^2} \int_0^{\infty} x d x J_3^2(k \lambda x) \frac{x^2-2}{(x^2+1)^{5/2}} \nonumber \\
& \times \int_x^{\infty} y d y J_3 (k \lambda y) N_3 (k \lambda y) \frac{y^2-2}{(y^2+1)^{5/2}},
\end{align}
with $N_3$ being the Neumann function.

\section {Order parameter and transition temperature} \label{TT}

In the regime of a weak interlayer attractive interaction, we use the BCS approach and obtain the gap equation for the momentum-space order parameter 
\begin{align} \label{gap}
\Delta ({\bf k})&= - \int \frac{d^2k'}{(2\pi)^2} f_3({\bf k'},{\bf k})\Delta ({\bf k'}) \\
& \times  \left[\frac{ \text{tanh}\left(\varepsilon_{k'}/2T \right)} {2\varepsilon_{k'}}-\frac{1}{2(E_{k'}-E_k-i0)}\right], \nonumber 
\end{align}
where  $\varepsilon_k=\sqrt{(E_k-\mu)^2+|\Delta (k)|^2}$ is the gapped dispersion relation, 
$\mu$ is the chemical potential and $E_k=\hbar^2k^2/2m$ is the energy of free particle. 
It is convenient to work with the gap equation projected onto the $f$-waves symmetry.
To this end, we use the expansions $\Delta ({\bf k})=\sum_l \Delta (k) e^{i \phi_k l}$ and $f_l ({\bf k},{\bf k'})=e^{i (\phi_k-\phi_{k'}) l} f_l (k,k')$ \cite{Fedo, Boudj, Zhen}.
Then, multiplying both sides by $e^{-i \phi_k l}$ and integrating over $d\phi_{k}$ and $d\phi_{k'}$, we find
\begin{align} \label{gap1}
\Delta (k)&= - {\cal P}\int \frac{d^2k'}{(2\pi)^2} f_3(k', k)\Delta (k') \\
& \times  \left[\frac{ \text{tanh}\left(\varepsilon_{k'}/2T \right)} {2\varepsilon_{k'}}-\frac{1}{2(E_{k'}-E_k-i0)}\right], \nonumber 
\end{align}
where ${\cal P}$ is the principal value of the integral.
Equation (\ref{gap1}) may be used for calculating $\Delta (k)$ and the superfluid transition temperature $T_c$. 

Setting apart the main contribution to the integral on the right-hand side of Eq. (\ref{gap1}) which arises from a narrow vicinity of the Fermi surface,
we obtain a relation between $\Delta (k)$ and $\Delta (k_F)$ as
\begin{equation} \label{gap2}
\Delta (k)= \Delta (k_F)  \frac{\bar f_3 (k_F,k)} {\bar f_3 (k_F)}.
\end{equation}
Employing Eq.(\ref{gap1}) makes it possible to obtain a relation between the zero-temperature order parameter on the Fermi surface $\Delta_0 (k_F)$
and $T_c$.

\begin{figure}
\centerline{
\includegraphics[scale=.8,clip]{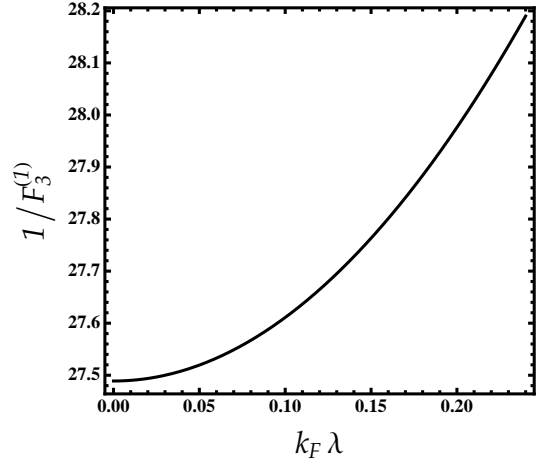}}
 \caption { Scattering amplitude $1/{\cal F}_3^{(1)}$ as a function of $k _F\lambda$ for $r_*/\lambda=9$.}
\label{SA} 
\end{figure}

Next we calculate the critical temperature by explicitly solving the gap equation (\ref{gap1}) at $k=k_F$. 
We split the region of integration into two sectors:$ |E_k-E_F| < \omega $ and $ |E_k-E_F| > \omega$. 
In the first sector the main contribution to $\Delta (k)$ comes from $k'$ close to $k_F$.
In the second part, $ |E_k-E_F| > \omega$, we set $\text{tanh}\left(\varepsilon_{k'}/2T \right)/ (2\varepsilon_{k'}) = 1/(2|E_{k'}-\mu|)$
and  keeping only the leading low-momentum contribution to the off-shell amplitude $\bar f_3 (k'_F,k)$ (for more details see Ref.\cite{Fedo, Boudj}). 
Then, summing all the contributions by omitting $\Delta (k_F)$ and putting $E_F \sim \omega$ in the terms proportional to $(k_F r_*)^2$, we obtain 
\begin{align} \label{gap3}
1= &(k_F r_*) {\cal F}_3^{(1)} (k_F\lambda ) \ln\left( \frac{2e^{\gamma -h_3 (k_F\lambda )}} {\pi} \frac{E_F}{T_c} \right)  \nonumber \\
& - (k_F r_*)^2 {\cal F}_3^{(2)} (k_F\lambda ) \ln\left( \frac{E_F}{T_c} \right), 
\end{align}
where $\gamma$=0.5772 is the Euler constant and the function $h_3$ is defined as
$$h_3(k_F\lambda )=-2\int_0^1 \frac{xdx}{1-x^2} \bigg\{\left[ \frac {{\cal F}_3^{(1)} (k_F\lambda, k_F\lambda x)} {{\cal F}_3^{(1)} (k_F\lambda)} \right]^2-1\bigg\}.$$
The effective mass has to be replaced instead of the bare mass in the gap equation  using the Fermi liquid theory \cite{Zhen}
\begin{equation}\label{EfM}
m^*=\frac{m}{1+\frac{4 k_Fr_*}{3\pi}}.
\end{equation}
In the limit $k_F r_* \ll1$,  the critical temperature is given as 
\begin{align} \label{Tcrt}
\frac{T_c}{E_F}&= \frac{2 e^{\gamma} }{\pi}  \exp \left [-h_3(k_F\lambda ) - \frac{ {\cal F}_3^{(2)} (k_F\lambda )}{{\cal F}_3^{(1)2}(k_F\lambda ) }  \right]  \\
 &\times \exp \left[- \frac{4}{35\pi {\cal F}_3 ^{(1)}(k_F\lambda)} -\frac{1}{k_F\lambda (r_*/\lambda) {\cal F}_3^{(1)}(k_F\lambda ) }\right] \nonumber\\
&=\frac{2 e^{\gamma} }{\pi} A_3 (k_F\lambda ).\nonumber
\end{align} 
Equation (\ref {Tcrt}) shows that the $f$-wave critical temperature decrease very rapidly for $k_F\lambda > 1.2$ due to the fast decay of the scattering amplitude $F_3^{(1)}$ (see Fig.\ref{SA}).
The optimal value of $k_F\lambda$ is around 0.9 with $T_c$ reaching values on the order of $\sim 8 \times 10^{-12} E_F$. 

Let us now look how this novel $f$-wave superfluid of fermionic polar molecules is experimental feasible.
For ${}^{40}$K${}^{87}$Rb molecules with a dipole moment of $d=0.566$ D and $r_*=6 \times 10^3 a_0$  ($a_0$ is the Bohr radii) \cite{Jin} 
at the density $n \sim 7 \times 10^9$ cm$^{-2}$ in each layer, one has an interlayer separation $\lambda =30$ nm for $k_F \lambda=0.9$.  
In this case,  the critical temperature is around $T_c \sim 20$ fK indicating that molecules reach a new record temperature, colder than what has been attained before.
Strictly speaking, the critical temperature is related to the density which should be large enough to get higher values for $T_c$.
For instance, for ${}^{87}$Rb${}^{133}$Cs molecules ($d=1.3$ D \cite{Jin}) with density $n\sim 10^{11}$  cm$^{-2}$, the $T_c$ 
is around one order of magnitude larger than the case of ${}^{40}$K${}^{87}$Rb, reaching $\sim 245$ fK.

Indeed, temperatures in the femtokelvin regime are not available in current experiences and remain challenging.
One can expect that such extreme low temperatures could be achieved by preparing polar molecules ensembles under microgravity \cite{Zoest, Mun, Rud}. 
This is because of lowering the trapping potential for the gas adiabatically without the need of levitation fields to compensate
for gravity. In microgravity, the lifetime for the free and undisturbed evolution of the condensate can be significantly enlarged \cite{Rain}.
Recently, microgravity has attracted considerable attention and has become a hot topic in the field of ultracold gases. For instance, 
European Space Agency, German Space Agency  and National Aeronautics and Space Administration (NASA) have launched a variety of programmes on ultracold atoms physics
in order to better understand  how complexity arises in the universe and to test fundamental laws of physics as well as to demonstrate an atomic
Clock Ensemble in Space.


\section{conclusion} \label{Conc}

In summary, we have shown the occurrence of novel $f$-wave superfluids of fermionic polar molecules confined into a bilayer setup
with dipole moments are in opposite directions in different layers at temperatures of the order of few femtokelvins.
These exotic states, which, to our knowledge, have not been unambiguously
identified before in cold atom systems, thus our results will greatly enrich the investigation of unconventional superconductors and superfluids.
Obtaining such pairing states  would pave the way  to a better understanding the phenomenon of  exotically  topological $f+i f$ phase, 
quantum interferometry, quantum metrology, Majorana modes, non-Abelian statistics and topologically protected quantum information science.

\section{Acknowledgements}
Stimulating discussions with Gora Shlyapnikov, Wolfgang Ketterle,  and Jun Ye  are acknowledged.

\end{document}